\documentclass[12pt,a4paper]{article}
\usepackage[ascii]{inputenc}
\usepackage{color, colortbl}
\usepackage{amsmath}
\usepackage{booktabs}
\usepackage{amsfonts}
\usepackage{amssymb}
\usepackage{tabularx}
\usepackage{makeidx}
\usepackage{graphicx}
\usepackage{natbib}
\usepackage{lineno}
\usepackage[title]{appendix}
\usepackage{authblk}
\usepackage{hyperref}
\usepackage{multirow}
\newcommand{\comment}[1]{}

 \hypersetup{
     colorlinks=true,
     linkcolor=blue,
     filecolor=blue,
     citecolor = blue,      
     urlcolor=cyan,
     }
\usepackage{arydshln}
\usepackage[flushleft]{threeparttable}
 \usepackage{natbib}
\usepackage{subfig}
\usepackage{float}
\usepackage{subfloat}
\usepackage[svgnames]{xcolor}
\usepackage{listings}
\definecolor{Gray}{gray}{0.9}
\date{}
\usepackage[width=17.00cm, height=23.00cm]{geometry}
\usepackage[flushleft]{threeparttable}

\author[1,2*]{Osafu Augustine Egbon}
\author[1,3]{Asrat Mekonnen Belachew}
\author[4]{Ezra Gayawan}
\author[1]{Francisco Louzada}
\affil[1]{\small Institute of Mathematical and Computer Sciences, University of S\~ao Paulo, S\~ao Carlos, Brazil}
\affil[2]{\small Department of Statistics, Universidade Federal de S\~{a}o Carlos, S\~{a}o Carlos, Brazil}
\affil[3]{\small Department of Mathematics, Ambo University, Ambo, Ethiopia}
\affil[4]{\small Department of Statistics, Federal University of Technology, Akure, Nigeria}
\affil[*]{\small Corresponding Author: Osafu Augustine Egbon (eosafu.a@gmail.com)}

\title{Violent event-related fatality patterns in Ethiopia: a Bayesian spatiotemporal perspective}
\begin{document}
\maketitle
\begin{abstract}
Fatalities resulting from violence in armed conflict have long been a significant public health issue in Ethiopia. Despite the severity of this problem, more comprehensive quantitative scientific studies need to be conducted to elucidate the sequence and dynamics of these occurrences. In response, this study introduces a spatio-temporal statistical method designed to uncover the patterns of fatalities associated with violent events in Ethiopia. The research employs a two-part zero-inflated Bayesian generalized additive mixed model, which integrates a spatio-temporal component to map the fatality patterns across Ethiopian regions. The dataset utilized originates from the Armed Conflict Location and Event Data Project, covering fatality counts related to violent events from 1997 to 2022. The analysis revealed that nine out of thirteen administrative regions exhibited a probability greater than 0.6 for fatality occurrence due to violent events, with five regions surpassing a 0.7 probability threshold. These five regions include Benishangul Gumz, Gambela, Oromia, Somali, and the South West Ethiopian People's Region. Notably, the Tigray region displayed the highest probability (0.558) of experiencing more than 20 deaths per violent event, followed by the Benishangul Gumz region with a probability of 0.306. Encouragingly, the findings also indicate an average decline in fatalities per violent event over time. Specifically, the probability of more than 20 deaths per event was 0.401 in 2020, which decreased to 0.148 by 2022. These insights are invaluable for the government, policymakers, political leaders, and traditional or religious authorities in Ethiopia enabling them to make informed, strategic decisions to mitigate and ultimately prevent violence-related fatalities in the country.
\end{abstract}
\newpage



\subsection*{Ethics Declaration}
Not applicable.



\section{Introduction}
Ethiopia has experienced several civil conflicts and wars with neighbouring countries for decades. The consistent conflicts in the region could be linked to several years of political instability and tension \citep{opalo2021ideology}. 
Violence in Ethiopia frequently arises as a result of political instability, social disorder, economic challenges, and inter-ethnic conflicts within the region. The surge in violent events since 2017 is mainly due to the rise in militant groups, different competing ethnonationalism in the context of state fragility, and party institutions \citep{yusuf2019drivers}. In recent years, Ethiopia has witnessed frightful violent activities, which is a significant burden to social and public health in the state. Ethnic conflicts in the region nowadays are a major problem, characterized by public unrest, especially since 2016. Such events feature the destruction of infrastructure, loss of agricultural resources, forced internal displacement of millions of people, and ultimately, deaths. As a consequence, there are severe deprivations of good education, health,  and food security \citep{devereux2000food,gonfa2019determinants}.

Despite Ethiopia's impressive economic performance over the past decade, the recent decline has posed a formidable challenge for both policymakers and the general population. Conflicts in Ethiopia have had a detrimental impact on access to agricultural products and raw minerals, which are crucial drivers of Ethiopia's economic development \citep{bekana2022regression}. This, in turn, could further exacerbate the current situation in the country. For example, the conflict in the Tigray region alone cost the Ethiopian government over one billion dollars by 2021, and this figure may rise if the conflict persists \citep{blanchard2021ethiopia}. These conflicts not only diminish Ethiopia's economic capacity but also present a grave economic and security threat to neighbouring countries, particularly in terms of the influx of refugees.

According to the Global Organized Crime Index \citep{WOCI}, violent events such as massacres, human smuggling, and human trafficking increased in Ethiopia and are above the world's average.  This has primarily been fueled by the illicit inflow of weapons in Ethiopia from neighboring countries, as well as the military forces in Somalia \citep{tilahun2020financial}. In 2021, Ethiopia reported 5.1 million internally displaced people within twelve months, which is the highest per year in the world \citep{CPA2022}. 
Ethiopia has experienced several conflicts. For instance, since 1998, the Ogaden National Liberation Front (ONLF) and the Ethiopian Defense Force have been at war in Somalia, demanding autonomy for Ethiopia's Somali region \citep{abdullahi2007ogaden}. Since 1973, when the Oromo Liberation Front (OLF) attempted to create an independent state of Oromia, there has been violence in Oromia between the OLF and the Ethiopian government \citep{baxter1996being}. Beyond conflict with governmental forces, inter-ethnic conflicts are also prevalent in Ethiopia. For example, ethnic Somali and Oromos armed clash in the Oromia region over a disagreement on grazing lands \citep{PP2022}, that may have resulted from the impact of climate changes. The conflict led to violence against civilians. For instance, according to the United Nations Office for the Coordination of Humanitarian Affairs, hundreds of people were killed and over 1.5 million people left their homes in the year 2018 \citep{bayu2021conflict}. Moreover, Ethiopia has had armed territorial conflicts with neighboring countries, such as Eritrea, Sudan, and Kenya \citep{kebede2020impacts}.

Understanding the pattern of conflict, violence, and the resulting fatalities is crucial for developing a realistic model for mitigating the impact of these armed activities on the Ethiopian populace and economy. The causes, nature, and effects of city crimes in Ethiopia have been the subject of various scientific research studies, however, there are no actual quantitative studies on conflict, violence, and fatalities in Ethiopia. For instance, \cite{desalegn2011ordinal} used the ordinal logistic regression model and identified that the major factors that affect crime outcomes are socio-demographic factors such as sex, age, education status, marital and employment status, and environmental factors such as urban, rural by using reported data from Tigray region, Ethiopia. Similarly, \cite{hateu2016assessment} showed that social and economic factors affect participants' propensity to commit crimes; nevertheless, certain conditions, poor socialization, parental negligence in providing sufficient supervision for the children, peer pressure, and a lack of resources are few examples of factors that strongly influenced participants' potential to commit a crime. \cite{kassa2020prevalence} conducted a meta-analysis on violence against women in Ethiopia and found that physical, sexual, and psychological violence against women in the country is on a substantial level.
Using the transactive approach, \cite{yeshiwas2018transactive} found that ethnic conflict in Ethiopia arises as a result of feelings and thoughts developed by members of a particular group as they engage and correlate their positions with those of other ethnic origins. \cite{teferra2018substance} and \cite{fekadu2007status} identified that, among high school and college students in Ethiopia, there is a high degree of drug use, alcohol, khat chewing, and cigarette smoking,
family unemployment, family conflict, and depression, which are some of the elements elites of armed groups use as a yardstick to recruit young and vibrant students to carry out violent events.

The earlier aforementioned scientific studies were only concerned with categorizing the different types of crimes and the effects that criminal and violent incidents had on the country's sub-regions. Moreover, the occurrence of violence in conflict or non-conflict events exhibits some spatial characteristics, that have been neglected by previous studies. It is evident that locations close by are more likely to experience similar patterns of violent events compared with distant ones. Hence, the role of spatial dependence cannot be neglected. Therefore, this research unveils the spatio-temporal pattern of fatalities arising from violence in armed conflict and non-conflict events in Ethiopia, which is relevant for understanding the progression of violent activities and their related burden to the nation. 

{The acquired fatality data resulting from violent events contain an excessive number of zeroes, which could introduce bias in modeling. To address this issue, we employed a two-part zero-inflated Bayesian generalized additive mixed model with spatio-temporal effects on a continuous spatial scale. This model allows for analyzing the spatio-temporal pattern of fatalities on the probability of fatality occurrence and fatality counts. The adopted framework was based on the work proposed in \cite{krainski2018advanced}, \cite{sadykova2017bayesian}, and \cite{asmarian2019bayesian}. Though these frameworks are elegant, they suffer computational limitations, especially when modeling a continuous spatio-temporal pattern as in this work. To overcome this challenge, we implemented a sequential computation approach for the binary and count components of the zero-inflated model, which essentially divides the computational burden into smaller chunks. The main difference between our approach and previous works is that; first, our approach provides a means for distributing computational burden into smaller chunks, and second, we provide a framework for modeling continuous spatio-temporal patterns. We utilized {a Stochastic Partial} Differential Equation to model the continuous spatio-temporal pattern to enable predictions about the likelihood of fatality patterns in regions where data have not been observed. By examining the results, we aim to gain insights into the patterns of fatalities from violent events and quantify the impact of various risk factors. This information can be valuable to policymakers, as well as for those involved in policing, defense forces, and resource management and distribution in Ethiopia.}



\section{Materials and Methods}\label{Method}
\subsection{Data}
The data used in this study was obtained from the \href{www.acleddata.com}{\textit{Armed Conflict Location and Event Data Project (ACLED)}}. It is a project that collects data on protest, crime, and violent events and provides the event type, actors, year, locations, and fatalities reported across the globe. The records of violent-related events in Ethiopia were first extracted on October 8, 2021, and updated on July 18, 2022, {across all regions of Ethiopia namely: {Addis Ababa, Afar, Amhara, Beneshangul Gumuz, Dire Dawa, Gambela, Harari, Oromia, Sidama, SNNPR, Somali, South West, and Tigray}. The data contain violent-related events from the year 1997 to 2022. Over the years, $8,490$ events were recorded. {Figure 1 presents the spatio-temporal plot of the natural logarithm of the observed fatality counts during the study period, where one was added to the count to cater for the zero cases when taking the logarithm.}
In this study, the response variable of interest is the fatality counts, and the covariates are the event type, temperate season, year, and spatial locations where the events occurred. Additionally, a categorical covariate was created from the notes corresponding to each event as a way to measure the impact of the activities of organized groups on fatality. The variable recorded the names of the social, political, or armed groups associated with the violent event. The groups extracted include the Eritrea army, Ethiopian Defense Force (EDF), Ethiopian Unity and Freedom Force (EUFF), Ginbot7, Ogaden National Liberation Front (ONLF), Oromo Liberation Army (OLA), South Sudan People's Defence Forces (SPLA), Tigray People's Democratic Movement (TPDM), and  Tigray People's Liberation Front (TPLF). These covariates are tabulated in Table \ref{tabcov}.

\begin{table}[H]
    \centering
\caption{Summary of variables included in the model.}
\resizebox{\textwidth}{!}{    \begin{tabular}{llll}
   & Variable   & Type&Values \\
         \hline
Response  &  Fatality count & Numeric &min $= 0$, median $= 1$, max $=1172$.\\
&&&\\
   \rowcolor{Gray}
\multirow{1}{*}{Covariate} & Event type&Categorical &Violent demonstration, Suicide bomb, Shelling, Sexual violence, Remote explosive,\\
   \rowcolor{Gray}
&&&Government regains territory,  Attack, Armed clash,  Abduction/forced disappearance,\\
   \rowcolor{Gray}
&&& Non-state actor overtakes territory, Mob violence, Grenade, Air/drone strike.\\
& Temperate season& Categorical &Summer, Winter, Spring, and Autumn\\
   \rowcolor{Gray}
   &Groups&Categorical&Eritrea army, EUFF, Ginbot7, ONLF, OLA, SPLA, TPDM, and TPLF.\\
&Year&Numeric&1997, 1998,...,2022.\\
   \rowcolor{Gray}
&Location&Numeric& Longitude, Latitude.\\
\hline
    \end{tabular}}
    \label{tabcov}
\end{table}
\begin{figure}
    \centering
    \includegraphics[scale=0.5]{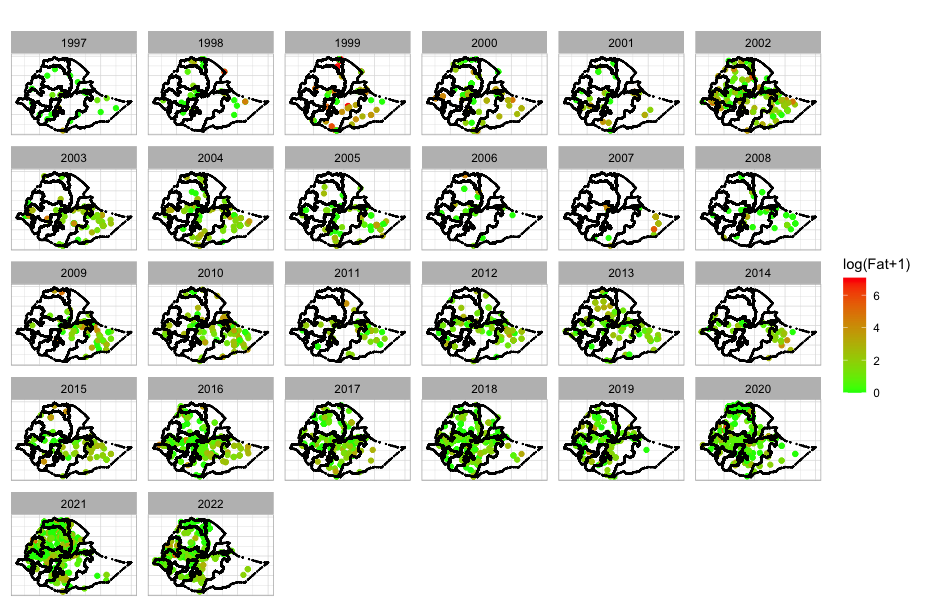}
    \caption{The logarithm of the observed fatality counts plus 1 from 1997 to 2022.}
    \label{fig:spatdescript}
\end{figure}
Since we aim to monitor fatality across the years in Ethiopia, a population variable was created and used as offset since the occurrence of these events has a strong positive relationship with the population size \citep{quick2019multiscale}. The population data were obtained from the World Bank database, openly available at \href{https://data.worldbank.org/indicator/SP.POP.TOTL}{https://data.worldbank.org}.  
\subsection{Statistical method}
\subsubsection{Response distribution}
Let $Y(s), s\in \mathcal D$ be a random variable representing the number of fatalities at location $s$, and $\mathcal D$ be the territory of Ethiopia. At any $s\in \mathcal D$, we assume that $Y(s)$  can be represented by some probability distribution $P(\Lambda(s))$ conditioned on some latent process through the parameter $\Lambda(s)$. This implies that the number of fatalities per event varies within $\mathcal D$. The statistical problem is to estimate the spatio-temporal latent process generating the fatality and make projections over the whole spatial domain $\mathcal D$ to understand the pattern of occurrence. A common choice for $P$ is the Poisson distribution, which has its mean equals its variance. If the data is drawn from a highly heterogeneous or clustered population, there would be greater conditional variance than the one assumed by the Poisson distribution \citep{berk2008overdispersion}. Thus, overdispersion would occur assuming the Poisson model for such data, causing the model to underperform \citep{cox1983some}. A competing choice for $P$ is the Negative Binomial distribution which has been adopted in several works to overcome overdispersion in data \citep{ver2007quasi}. Another standard competing choice for $P$ is the generalized Poisson distribution \citep{joe2005generalized}, which has been shown to be useful for modeling count data with long right tails \citep{joe2005generalized}.

\begin{figure}[H]
    \centering
    \includegraphics[scale=0.7]{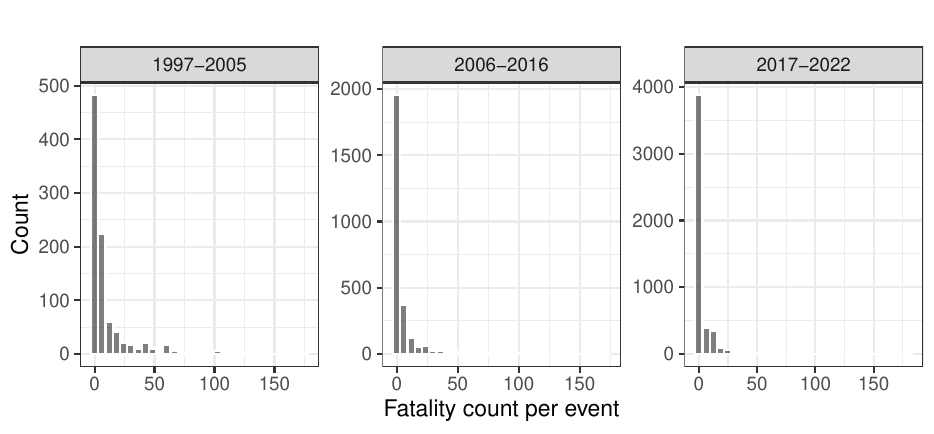}\\
    \caption{{Observed fatality counts} per violent event in Ethiopia.}
    \label{bar}
\end{figure}

Moreover, the fatality data acquired for this study is inflated with zeros. This is evident from the bar plot in Figure \ref{bar} for intervals of years. The Figure shows that {the frequencies of zeros are relatively very high compared with nonzeros, which may indicate that the zeroes} may exceed the expected zeros under a standard count model, a problem that could make a standard count probability distribution for $P$ to underperform {(see Figure S1 in the supplementary material for an example)}.  {An intuitive example is that in Ethiopia, there are safe regions where fatalities are rare and zero fatalities are recorded, and there are high-risk regions where fatalities are common but zero fatalities are also sometimes recorded. This leads to zero inflation in the dataset, as these two types of zeros are generated by different processes. }
We adopted a zero-inflated model to account for the data imbalance and control for overdispersion \citep{zuur2009zero}. The probability distribution follows as
\begin{align}
   \mathcal P(Y(s)=y(s)\mid \Lambda(s))= (1-\pi(s)) \mathbb I_{(y(s)=0)} + \pi(s) P(y(s)\mid \Lambda(s)),
   \label{zi}
\end{align}
where $\pi(s)$ is the probability of fatality at $s$ and $\mathbb I_{(a)}$ is an indicator function that is one only if $a$ is true. 
The acquired data portrays the possibility of variation in the zero fatality occurrence {across time} as shown in Figure \ref{bar}. A possible way to ensure that such variation is captured is to model $\pi(s)$ using a probabilistic model. Therefore, we consider linking  {$\pi(s)$} to covariates. 

{
Several zero-inflated models within a Bayesian framework have been proposed in the literature to model binary and count components based on covariates, and utilizing \texttt{R-INLA} tool for inference \citep{sadykova2017bayesian,krainski2018advanced,asmarian2019bayesian}. For instance, Asmarian et al. (2019) developed a joint model using \texttt{R-INLA} to investigate the prevalence of male cancer in Iran. While the joint model is elegant, it can be computationally demanding, particularly in spatio-temporal modeling scenarios with a large volume of data points, and may not adequately propagate uncertainty between both model components. In the acquired data, the joint modeling approach using \texttt{R-INLA}, as proposed by \cite{asmarian2019bayesian}, required 96 hours to complete estimation on a computer equipped with 12 cores and 32GB of RAM. Estimation was unattainable on computers with lower computational capabilities unless model precision was sacrificed through mesh triangulation.
} To avoid {such} computational {challenge}, one may consider approaching the modeling of spatio-temporal patterns in the zero and count parts of model \ref{zi} using a {sequential} modeling technique, which decomposes the computation burden into smaller chunks. 

We decomposed the zero-inflated model {into binary and count components. Let $Z_{0i}$ be a binary random variable in the binary component and $Z_{1i}$ be a count variable in the count component.} We defined the observed $Z_{0i}$ as
\begin{align}
    z_{0i}=
    \begin{cases}
        1 &\text{ if }y(s_i) > 0\\
        0&\text{ otherwise}.
    \end{cases}
\end{align}
{$Z_0$ quantifies the occurrence or non-occurrence of fatality for every violent event that occurs}. It is modeled as

\begin{align}
\begin{aligned}
    Z_{0i}\sim &\,\text{Bernoulli} (\pi_i),\\
       \pi_i =& \frac{\exp\Big(\eta^{(b)}_i\Big)}{1+\exp\Big(\eta^{(b)}_i\Big)},\forall s_i,t_i, i =1,2,...,n,\\
  \end{aligned}
\end{align}
{where $\eta^{(b)}_i\in \mathbb R$ is a structured linear predictor for the binary component defined in Equation \eqref{LinearPred1} or \eqref{LinearPred2} below. $n$ is the total number of violent events over the years and $s_i$ is the spatial location where the $i$th violent event took place in year $t_i$. {That is, for every event $i$, there is an associated spatial location $s_i$ and year $t_i$}}. $\pi(s)$ can be interpreted as the probability of fatality at location $s$.

We define the observed count variable $Z_1$ as
\begin{align}
    z_{1i}=
    \begin{cases}
        \text{NA} & {\text{ if } y(s_i) = 0}\text { and }  {\Tilde\pi}_i<c,\\
        0& {\text{ if } y(s_i) = 0 \text { and }   {\Tilde\pi}_i\geq c},\\
        y(s_i)& \text{ otherwise}.
    \end{cases}
\end{align}
{where $c\in(0,1)$ is a threshold and $  {\Tilde\pi}_i = P(Z_{0i}=1| Data)$ is the posterior predictive probability of the binary component equals one at location $s_i$ and $Data$ contains both covariates and response used in the modeling of the binary component. NA indicates that $y(s_i)=0$ is a structured zero, and when $z_{1i} = \text{NA}$, it implies that $y(s_i)$ does not contribute to the likelihood of the count component. Here, we have assumed that the covariates contain information that can differentiate structured zero from unstructured zero (zero count). This is appropriate because covariates such as ``violent event type", ``temperate season" and ``location" play important roles in fatality occurrence and counts. For a given threshold $c$, $ {\Tilde\pi}_i\geq c$ implies that $y(s_i)=0$ is an unstructured zero (zero counts), otherwise, it is a structured zero. This means that data points with high $ {\Tilde\pi}_i$ have low $1- {\Tilde\pi}_i = P(Z_{0i}=0| Data)$, and are more
likely to be zero count. The two special cases are $c=0$ and $c=1$. $ c=0$ implies that all $y(s_i)=0$ are count zero and $c=1$ means that all $y(s_i)=0$ are structured zero.  A cross-validation technique is used to choose the appropriate $c$ that results in the smallest WAIC \citep{gelman2014understanding} {evaluated only by the non-zero part of the data used in the modeling due to variation in the sample size for different values of c}. {The stochastic representation of the model is given in the supplementary material.} } 


 We compared the performance of different models on $Z_1$. Specifically, we assumed $Z_{1i}\sim Poisson(\lambda_i$), $Z_{1i}\sim NegBinomial(\mu_i,\xi)$, and  $Z_{1i}\sim GPoisson (\phi_i,\psi,p)$ as competing models, {where $NegBinomial$ and $GPoisson$  are the negative Binomial and generalized Poisson distributions respectively.} The reparameterized probability distributions  adopted follow as

{\footnotesize
\begin{align*}
    Z_{1i}\sim Poisson(\lambda_i) &=\frac{\lambda_i^{z_{1i}}\exp(-\lambda_i)}{z_{1i}!},\, \text{where }{\lambda_i=\exp\Big(\eta^{(c)}_i\Big)},\\
 Z_{1i}\sim NegBinomial(\mu_i,\xi)&= \frac{\Gamma(z_{1i}+\xi)}{\Gamma(\xi)z_{1i}!} \mu_i^{z_{1i}}(1-\mu_i)^{\xi},\,\text{ where } {\mu_i= \frac{\exp\Big(  \eta^{(c)}_i\Big)}{1+\exp\Big(\eta^{(c)}_i\Big)}},\\
 Z_{1i}\sim GPoisson(\phi_i,\varphi,p)&=\frac{\phi_i(\phi_i+\varphi\phi_i^{p-1} z_{1i})^{z_{1i}-1}}{\left(1+\varphi\phi_i^{p-1}\right)^{z_{1i}}z_{1i}!}\exp\Big(-\frac{\phi_i+\varphi\phi_i^{p-1}z_{1i} }{1+\varphi\phi_i^{p-1}}\Big), \text{where}\\
 {\phi_i}&{= \exp\Big(\eta^{(c)}_i\Big)},\, 
\end{align*}
}
{\noindent where $\eta^{(c)}_i\in\mathbb R$ is a structured linear predictor for the count component defined in Equation \eqref{LinearPred1} or \eqref{LinearPred2} below. The mean of the Poisson, negative binomial, and generalized Poisson distributions are respectively $\lambda_i,$ $\xi\exp(\eta^{(c)}_i)$, and $\phi_i$, and their corresponding variances are $\lambda_i,$ $\xi\exp(\eta^{(c)}_i(1+\exp(\eta^{(c)}_i))$, and $\phi_i(1+\varphi\phi_i^{p-1})^2$ respectively}. Particularly in this work, $p=1$. 

The benefit of this approach is that the parameters that drive the binary and count components can be estimated sequentially, which relatively has a lower computation burden compared with the joint analysis proposed in \cite{asmarian2019bayesian}. {The framework involves first estimating the binary component and determining $ \hat{\Tilde\pi}_i$  and $c$. Conditioning on $ \hat{\Tilde\pi}_i$  and $c$, the count component is estimated}. This is advantageous, especially in situations where a smooth spatio-temporal pattern over spatial domain $\mathcal D$, which is usually computationally intensive, is to be obtained. The {main difference between the above specification and that of \cite{asmarian2019bayesian} is that; first, our approach provides a means for distributing computational burden into smaller chunks, which is useful for processing complex models with relatively lower computational power. Secondly, it provides a framework for modeling continuous spatio-temporal patterns.} 

\subsubsection{Structural linear predictors}

\cite{berk2008overdispersion} argued that the challenge of overdispersion can be alleviated using the negative Binomial and Generalized Poisson only if the problem of lack of fit of the Poisson distribution on overdispersed data does not stem from the stochastic part of the model. That is, excess variation arising from the wrong systematic structure of the model cannot be corrected using other distributions. Hence, we adopted two structural forms for $\eta^{(b)}_i$ and $\eta^{(c)}_i$. {This section contains two subsections with the first presenting the 
specifications of the structural forms adopted while the second 
 defines the terms in the linear predictors.}

\textbf{{\footnotesize {Specifications of the linear predictors}}}
{
\begin{itemize}
    \item[1.] Structural form I: we first considered a structural form where spatial and temporal terms in the linear predictors are separate in the binary and count components. That is

\begin{align}
\begin{aligned}
    \eta^{(b)}_i=&\,\mathbf x_i(s_i)\boldsymbol\beta_0+f_0(t_i)+u_0(s_i),\\
 \eta^{(c)}_i=&\,\log(E(s_i,t_i))+\mathbf x_i(s_i)\boldsymbol\beta_1+f_1(t_i)+u_1(s_i),\,i=1,2,...,n,\\
    \end{aligned}
    \label{LinearPred1}
\end{align}

\noindent where $\mathbf x_i(s_i)$ is a design vector used to introduce model intercepts and fixed covariates such as violent actor, violent event type, and the temperate season.  $\boldsymbol\beta_0$ and $\boldsymbol\beta_1$ are respectively the corresponding model effects in the zero and count components that quantify how the fixed covariates contribute to the probability of fatality occurrence and the number of fatalities per event. {$f_0$ and $f_1$ are nonlinear functions of time that quantify the nonlinear temporal effect on fatality, and $u_0(.)$ and $u_1(.)$ are the spatial effects that quantify the spatial patterns of fatalities in both model components.  The offset variable $E(s_i,t_i)$ is the population of the administrative region where location $s_i$ belongs at time $t$. For example, if $s_i$ is a location within the Afar region, then $E(s_i,t_i)$ is the total population in Afar at year $t_i$.}

\item[2] Structural form II: according to \cite{blangiardo2013spatial}, it is important to account for space and time interaction which would explain time and space risk surface changes. Hence, we introduce a spatio-temporal term in the linear predictor. That is

\begin{align}
\begin{aligned}
    \eta^{(b)}_i&=\mathbf x(s_i)\boldsymbol\beta_0+  v_0(s_i,t_i),\\
     \eta^{(c)}_i&=\log(E(s_i,t_i))+\mathbf x(s_i)\boldsymbol\beta_1+v_1(s_i,t_i),\,i=1,2,...,n,
\end{aligned}
\label{LinearPred2}
\end{align}
where $v_0(s_i,t_i)$  and $v_1(s_i,t_i)$ are spatio-temporal effects for the binary and count components respectively.


 \end{itemize}}
 {Notice that structural form I considered both the spatial and temporal effects separately by assuming they are independent, whereas structural form II considered a spatio-temporal effect, by assuming an interrelationship between space and time. The linear predictors incorporated different interactions, which could be significantly relevant in understanding the pattern of fatality. We investigated these structural forms on how well they described the true structural form generating the data using the Deviance information criterion (DIC) and Watanabe-Akaike Information Criteria (WAIC). {In addition, we investigated a simpler baseline structural form, which disregards spatio-temporal interrelationships in the count component. That is,   
     $\eta^{(c)}_i=\log(E(s_i,t_i))+\mathbf x(s_i)\boldsymbol\beta_1$.  The baseline structure is important for benchmarking the two structural forms above. The next subsection 
     explains the terms $f_0,f_1,u_0,u_1,v_0$, and $v_1$ in the linear predictors.}
}

{\footnotesize\textbf{{Definition of terms in the linear predictors}}}

\noindent The nonlinear 
functions $f_0$ and $f_1$ were modeled with a cubic basis function defined as  
\begin{align}
    f_h(t_i) = \sum_{l=1}^{L} b_{hl}(t_i)\psi_{hl},\,h=0,1, 
    \label{spline}
\end{align}
where $b_{hl}(t)$ is a basis function, $\psi_l$ is the corresponding coefficient, and $L$ is the total number of knots. Details of the definitions of the basis functions are included in the supplementary material. The summation given in Equation \eqref{spline} can be written in matrix form {for $\mathbf t=(t_1,t_2,...,t_n)$} as $f_h(\mathbf t)= \mathbf B_h \boldsymbol\psi_h$, {where vector $f_h(\mathbf t)=(f_h(t_1),f_h(t_2),...,f_h(t_n))^T$}, and $\boldsymbol\psi_h=(\psi_{h1},\psi_{h2},...,\psi_{hL})^T$, and each row of the $n\times L$ basis matrix $\mathbf B$ corresponding to $t_i$ is $\mathbf b(t_i)= (b_1(t_i),b_2(t_i),...,b_l(t_i))$. {Hence, for the binary component, $f_0(\mathbf t) = \mathbf B_0\boldsymbol\psi_0$ and for count component $f_1(\mathbf t) = \mathbf B_1\boldsymbol\psi_1$}.
 The cubic spline has been shown to perform well in capturing nonlinear patterns \citep{crainiceanu2005bayesian}. 
 
 {For the spatial and spatio-temporal components of the linear predictor, } the intention is to estimate the function $u(s_i)$ and $v(s_i,t_i)$ using the observed data at $s_i\in \mathcal D, {\forall i}$ and then make projections across all locations and years using the estimated function. 
 A popular technique commonly adopted is the semi-variogram model \citep{jian1996semivariogram}, however, in this work, {we considered a Stochastic Partial Differential Equation (SPDE)} \citep{simpson2012think,egbon2024bayesian}.  The SPDE has been shown to perform adequately in many areas of application. 
 The SPDE model for any random field $d(s_i)\in\mathbb R$ is defined as 
\begin{align}
    (\kappa^2 - \Delta)^{(\alpha/2)} d(s_i)=\mathcal W(s_i),\, \alpha=\nu+1,\kappa>0,\nu>0,\forall i
\end{align}
whose solution is a {Gaussian Random Field (GRF)} with a Mat\'ern covariance function defined as 
\begin{align}
    {cov}(s_i,s_j) = \frac{\sigma^2}{2^{\nu-1}\Gamma(\nu)}\Big(\kappa\mid\mid s_i-s_j\mid\mid\Big)^\nu\mathcal K_\nu (\kappa \mid\mid s_i-s_j\mid\mid),
    \label{mat}
\end{align}
and the marginal variance $\sigma^2=\frac{\Gamma(\nu)}{\Gamma(\nu+1)4\pi\kappa^{2\nu}}$. {The GRF introduces spatial correlation, implying that fatality occurrences at proximate spatial locations are related. This ensures that fatality in one spatial location can be informative about the experience in nearby locations}. \cite{lindgren2011explicit} showed that the solution to the SPDE can be represented on a triangular mesh formed by a sequence of piecewise basis functions. That is $d(s_i)=\sum_{k=1}^{K}g_k(s_i)\alpha_k$, where $g_k(s_i)$ is the $k$th basis function evaluated at location $s_i$ and $\alpha_k$ is the corresponding parameter. {For all $\mathbf s= (s_1,s_2,...,s_n)$, the piecewise basis function can be written in matrix form as  $d(\mathbf s)= \mathbf G\boldsymbol\alpha$, where $ \mathbf G$  is {an $n\times K$} projection matrix with $(i,k)$ entry given as $g_k(s_i)$, and {$\boldsymbol\alpha=(\alpha_1,...,\alpha_K)^T$}. { $\boldsymbol\alpha$ is the parameter vector to be estimated from the data and the details of the derivation of $\mathbf G$ is found in \cite{lindgren2011explicit}.}}

{Specifically, the SPDE model was assumed for the spatial term} $u(s_i)$. Moreover, a similar model was assumed for the spatio-temporal term $v(s_i,t_i)$ but {grouped by unique time points. 
Likend to $d(\mathbf s)$, the SPDE  model for the {pure spatial} effect for the binary component is $u_0(\mathbf s)= \mathbf A\boldsymbol\theta_0$, where $\mathbf A$ is the projection matrix and $\boldsymbol\theta_0$ is the corresponding parameter.  For the binary {spatio-temporal} component, the SPDE model for the collection of all data points at year $t$ 
is $v_0(\mathbf s,t)= \mathbf C_t\boldsymbol\phi_{0t}$. That is $(\mathbf s,t)$ contains all the indexes of the data points of violent events in a given year $t$ across the country.  Similarly, for the count component, the spatial effect is $u_1(\mathbf s)= \mathbf A\boldsymbol\theta_1$ and the spatio-temporal effect is $v_1(\mathbf s,t)= \mathbf C_t\boldsymbol\phi_{1t}$, where $\mathbf C_t$ is the projection matrix at year $t$. The parameters $\boldsymbol\theta_0=(\theta_{01},...,\theta_{0K})^T$, $\boldsymbol\theta_1
=(\theta_{11},...,\theta_{1K})^T$,
$\boldsymbol\phi_{0t}=(\phi_{01t},\phi_{02t},...,\phi_{0Kt})^T$, and $\boldsymbol\phi_{1t}=(\phi_{11t},\phi_{12t},...,\phi_{1Kt}))^T$. Let $\boldsymbol\phi_{0}=(\boldsymbol\phi_{01}^T,\boldsymbol\phi_{02}^T,...,\boldsymbol\phi_{0T}^T)^T$, $\boldsymbol\phi_{1}=(\boldsymbol\phi_{11}^T,\boldsymbol\phi_{12}^T,...,\boldsymbol\phi_{1T}^T)^T$ and $\mathbf C$ be the corresponding block diagonal matrix of all the $\mathbf C_t, t=1,2,...,T$, where $T$ is the total number of years.}
Hence, the full hierarchical model, assuming structural form II for the binary component is given as
\begin{align}
\begin{aligned}
    Z_{0i}\sim &Bernoulli (\pi(s_i)),\\
       \pi_{i} =& \frac{\exp\Big(\eta^{(b)}_i\Big)}{1+\exp\Big(\eta^{(b)}_i\Big)},\forall s_i,t_i, i =1,2,...,n,\\
   \boldsymbol \eta^{(b)}=\,&\log(E(\mathbf s,\mathbf t))+\mathbf X(\mathbf s)\boldsymbol\beta_0
    + \mathbf C\boldsymbol\phi_0,
  \end{aligned}
\end{align}
where $ \boldsymbol\eta^{(b)}= ( \eta^{(b)}_1,...,\eta^{(b)}_n)^T$, $E(\mathbf s,\mathbf t)=(E(s_1, t_1),E(s_2, t_2),...,E(s_n, t_n))^T$, and $\mathbf X(\mathbf s)$ is a design matrix. The corresponding hierarchical model for the competing count models,
\begin{align}
\begin{aligned}
  Z_{1i}&\sim Poisson(\lambda_i), \,{\lambda_i=\exp\Big(\eta^{(c)}_i\Big)},\\
 Z_{1i}&\sim NegBinomial(\mu_i,\xi),\, {\mu_i= \frac{\exp\Big(  \eta^{(c)}_i\Big)}{1+\exp\Big(\eta^{(c)}_i\Big)}}, \text{ or}\\
 Z_{1i}&\sim GPoisson(\phi_i,\varphi,p),\,
 {\phi_i}{= \exp\Big(\eta^{(c)}_i\Big)},\,\\
   \boldsymbol \eta^{(c)}&=\,\log(E(\mathbf s,\mathbf t))+\mathbf X(\mathbf s)\boldsymbol\beta_1
    +\mathbf C\boldsymbol\phi_1,
    \end{aligned}
\end{align}
where $ \boldsymbol\eta^{(c)}= ( \eta^{(c)}_1,...,\eta^{(c)}_n)^T$. For the full hierarchical model using structural form I, it is sufficient to replace $\mathbf C\boldsymbol\phi_0$ with $\mathbf B_0\boldsymbol\psi_0+\mathbf A\boldsymbol\theta_0$ in  $ \boldsymbol\eta^{(b)}$ and $\mathbf C\boldsymbol\phi_1$ with $\mathbf B_1\boldsymbol\psi_1+\mathbf A\boldsymbol\theta_1$ in $ \boldsymbol\eta^{(c)}$.

\subsubsection{Bayesian inference}
In the frequentist approach, it is customary to penalize the log-likelihood using some penalization matrices. However, in the Bayesian framework, the uncertainty in the main effects is represented by a probability distribution. These distributions are referred to as the priors. In this section, we discuss the prior distributions for the main effects $\{\boldsymbol\beta, \boldsymbol\psi,\boldsymbol\theta, \boldsymbol\phi\}$, where $\boldsymbol\beta\in\{\boldsymbol\beta_0,\boldsymbol\beta_1\}$, $\boldsymbol\psi\in\{\boldsymbol\psi_0,\boldsymbol\psi_1\}$, $\boldsymbol\theta\in\{\boldsymbol\theta_0,\boldsymbol\theta_1\}$, and $\boldsymbol\phi\in\{\boldsymbol\phi_{0},\boldsymbol\phi_{1}\},$ and the corresponding hyper-parameters. 

According to literature, $\boldsymbol\beta$ is assigned a diffused Gaussian prior distribution centered at zero \citep{lemoine2019moving}. That is $\boldsymbol\beta\sim N(\mathbf 0,10^{4})$.
 A GMRF prior distribution centered at zero with a $K\times K$ precision matrix $\mathbf Q_{\theta}$ constructed based on the mesh triangulation was assumed for  $\boldsymbol\theta$. For technical details on how the covariance matrix is constructed for an SPDE model see \cite{lindgren2011explicit}. The prior distribution assumed for the spatio-temporal effect $\boldsymbol\phi = (\boldsymbol\phi_1,\boldsymbol\phi_2,...,\boldsymbol\phi_T)^T$, and each of the $\boldsymbol\phi_t=(\boldsymbol\phi_{1t},\boldsymbol\phi_{2t},...,\boldsymbol\phi_{Kt }),t=1,2,...,T$  is the GMRF modeled through a multivariate AR(1) process as follows:
 \begin{align}
    \boldsymbol\phi_t=\rho_\phi\boldsymbol\phi_{t-1}+\boldsymbol\epsilon_t,\,\boldsymbol\epsilon_t\sim N(\mathbf 0, \mathbf Q_\phi^{-1}),\;\mid \rho_\phi\mid<1,
\end{align}
 {where $ \mathbf Q_\phi$ and $ \mathbf Q_\theta$ are both obtained with the SPDE approach}. Assuming a weakly stationary AR(1) process, the marginal expectation $\mathbb E(\boldsymbol\phi_{t})=\mathbf 0$, variance Var$({\boldsymbol\phi_t})=\mathbf Q_\phi^{-1}(1-\rho_\phi^2)^{-1}$, and covariance Cov($\boldsymbol\phi_t,\boldsymbol\phi_{t^{\star}}$) $=\rho_\phi^{\mid t-t^{\star}\mid}$Var$({\boldsymbol\phi_t})$.  
For a given reference year, the AR(1) model indicates that fatality at a closer time behaves more similarly compared to those farther apart. {The spatio-temporal correlation $\rho_\phi$ quantifies the temporal dependence of fatality across years. A positive value of $\rho_\phi$ indicates that fatalities across the country are positively correlated over the years while a negative value of $\rho_\phi$ indicates that fatalities are negatively correlated over the years.} Similarly, from Equation \eqref{spline}, the effect $\boldsymbol\psi$ was modeled by a univariate AR(1) process
\begin{align}
    \psi_{l}=\rho_{\psi}\psi_{(l-1)}+\epsilon_{l},\,\,\epsilon_{l}\sim N(0, \mathbf \tau_\psi^{-1}),\;\mid \rho_{\psi}\mid<1,
\end{align}
where $l =1997,1998,...,2022$. The AR(1) model was assumed over higher order for computational reasons.  
All the assumed prior distributions apply to both the binary and the count components. Let $\mathcal X=\{\boldsymbol\beta, \boldsymbol\psi,\boldsymbol\theta, \boldsymbol\phi\}$  for instance, and the associated joint prior distribution is denoted by $h(\mathcal X)$ while the vector of hyperparameters is denoted by $\boldsymbol\omega$ with joint probability distribution $h(\boldsymbol\omega)$. Let $L()$ denote the likelihood function, then the joint posterior distribution follows as
\begin{subequations}
  \begin{align}
   \big[\mathcal X_0,\mathcal X_1,\boldsymbol\omega_0, \boldsymbol\omega_1\mid Data \big]&= \big[\mathcal X_1, \boldsymbol\omega_1\mid \mathcal X_0,\boldsymbol\omega_0,Data \big] [\mathcal X_0,\boldsymbol\omega_0\mid Data \big]\\
[\mathcal X_0,\boldsymbol\omega_0\mid Data \big] &\propto L_0(\mathbf z_0)h(\mathcal X_0)h(\boldsymbol\omega_0)\\
\big[\mathcal X_1, \boldsymbol\omega_1\mid \mathcal X_0,\boldsymbol\omega_0,Data \big]    &\propto L_1(\mathbf z_1\mid \Tilde{  \pi})h(\mathcal X_1)h(\boldsymbol\omega_1)
  \end{align}
      \label{post}
\end{subequations}
where $L_1(\mathbf z_1\mid \Tilde{ \pi})$ is the conditional likelihood of the count component, $\Tilde{ \pi}=(\Tilde{\pi}_1,\Tilde{\pi}_2,...,\Tilde{\pi}_n)$ and $\Tilde{\pi}_i = P(Z_{0i}=0| Data)$ is a vector of the posterior predictive probability of all the observed $\mathbf z_0$ in the binary component. Index $0$ denotes the binary component and index $1$ denotes the count component. Observed that parameters in the joint posterior distribution (Equation \eqref{post}a) {can be estimated sequentially by first estimating Equation \eqref{post}b, and then \eqref{post}c by conditioning on $\hat{\pi}$ derived from \eqref{post}b}. The posterior marginal distribution of the model parameters is used for inference.

\subsubsection{Posterior uncertainty}
The benefit of the Bayesian approach is the possibility of quantifying uncertainties in the posterior distribution. Let $[Z_0(s)\mid Data]$ represent the posterior predictive distribution in the zero component in location and $[Z_1(s)\mid Data]$ be the posterior predictive distribution in the count component in location $s$. We may be interested in quantifying the probability of fatality occurrence or the probability of a certain number of fatalities per violent event conditioned on the observed data. That is, we seek 
\begin{align}
\begin{aligned}
 P(Z_h(s)\in C_h \mid Data) &=  \mathbb E[\mathbb I_{(Z_h(s))\in C_h} \mid Data]\\
 &=\int\sum_{z_h(s)\in C_h}^{} [Z_h(s)=z_h(s),\mathcal X_h,\boldsymbol\omega_h \mid Data] d(\mathcal X_h,\boldsymbol\omega_h)\\
    &=\int\sum_{z_h(s)\in C_h}^{} [Z_h(s)=z_h(s)\mid\mathcal X_h,\boldsymbol\omega_h][\mathcal X_h,\boldsymbol\omega_h \mid Data] d(\mathcal X_h,\boldsymbol\omega_h),\\
    \end{aligned}
    \label{postpred}
\end{align}
where $h=0,1$, $\mathbb I$ is an indicator function, $[Z_h(s)\mid\mathcal X_h,\boldsymbol\omega_h]$ is the conditional likelihood, $[\mathcal X_h,\boldsymbol\omega_h \mid Data]$ is the posterior distribution given in Equation \eqref{post}, and $C_h$ is the set we intend to quantify its uncertainty. In this work, $C_0=\{1\}$ and $C_1=\{21,22,...\}$. In other words, we seek $P(Z_0(s)= 1\mid Data)$ and $P(Z_1(s)>20 \mid Data)$. {The threshold of $20$ was selected through data exploration. It represents the empirical average of observed counts across the country after excluding zeros}. The probability in the zero component measures the uncertainty of fatality occurrence, while the count component measures the uncertainty in fatality rates per violent event, {and the probability decreases with an increase in the threshold}. The integrals in Equation \eqref{postpred} are difficult to evaluate analytically. Instead, we evaluate it using Monte Carlo samples from the marginal posterior distributions. For example, draw sufficient large samples of the model parameters from their marginal posterior distributions. Substitute these samples into the corresponding probability distribution (Bernoulli or Negative Binomial) and draw samples from these probability distributions. Then estimate the proportions that belong to the corresponding set $C$.
\subsubsection{Estimation and Comparison with joint modeling}
We exploited the computational benefits of the \texttt{R-INLA} \citep{rue2009approximate} while leveraging its inbuilt capacity to handle missingness in the response variable. 
The model parameters were estimated within \texttt{R-INLA}, where the binary component is assigned ``\texttt{binomial}" and the count component assigned ``\texttt{gpoisson}", ``\texttt{poisson}" or ``\texttt{nbinomial}". We performed a simulation
study to compare the sequential approach adopted in this work and the joint modeling approach proposed in \cite{asmarian2019bayesian}.  The simulation results are described in the online supplementary material. Findings from the simulation study showed that both model presents equivalent results. 

{The $\texttt{R}$ code used for this analysis is accessible through {\href{https://github.com/eosafu/FatalityEthiopiaModel}{https://github.com/eosafu/Fatality}}} \href{https://github.com/eosafu/FatalityEthiopiaModel}{EthiopiaModel}.  \texttt{R-INLA} has become popular for Bayesian estimation as an alternative to the traditional MCMC estimation technique due to its good accuracy and relatively lower computational time \citep{held2010posterior}. It has been used for Bayesian estimation in various fields \citep{rue2009approximate,held2010posterior,egbon2024bayesian, egbon2025spatio}. Given the structure of the model presented here, it is easy to estimate the parameters in the zero and count components of the model using \texttt{R-INLA}. The \texttt{R-INLA} package incorporates the SPDE model and allows flexibility in the prior choice of the hyperparameters. In this case study, the Penalized complexity (PC) prior distribution \citep{fuglstad2019constructing} was assigned to the spatial range $r=\sqrt{8\nu}/\kappa,\,\nu=1$, and marginal variance $\sigma^2$, such that after sensitivity analysis, {$P(r< 1.42) = 0.9$ and $P(\sigma^2 > 1 ) = 0.9$} were chosen for both spatial and spatio-temporal effects. Moreover, $\tau_\psi$ and $\rho$ were assigned the PC prior distribution such that $P(\tau_\psi > 0.5)=0.9$, and $P(\rho_\phi > 0)=P(\rho_\psi > 0)=0.9$.

\section{Result}\label{result}
\subsubsection*{Descriptive statistics}
Figure \ref{fig:spaceplot} presents the descriptive statistics of violent event occurrences per $100,000$ population and the associated fatalities over the years and regions. The orange line in Figure \ref{fig:spaceplot}a, b, and c and the grey $95\%$ confidence intervals were obtained using local regression to show the temporal averages of events and fatalities throughout the study period. Specifically, Figure \ref{fig:spaceplot}a shows how the frequency of violent events evolves through time, and \ref{fig:spaceplot}d shows the corresponding descriptive spatial distribution. Based on the yearly estimated averages, the event rate per $100,000$ population showed a significant sharp increase 
 between the years $2015$ and $2022$, while the Oromia region had the highest event rate (Figure \ref{fig:spaceplot}d).  Figure \ref{fig:spaceplot}b shows the fatality rate per $100,000$ population over the years, and Figure \ref{fig:spaceplot}e shows the corresponding spatial distribution. Contrary to the event rates over the years, the average fatality rate declined consistently until $2019$, where it then picked up and kept increasing until the end of the study period. Moreover, in Figure \ref{fig:spaceplot}e, Tigray had the highest fatality rate per 100,000 population. Figure \ref{fig:spaceplot}c shows the fatality rate per violent event over the years, and Figure \ref{fig:spaceplot}f shows the corresponding spatial distribution. Overall, the rate decreased consistently throughout the study period, and Tigray is the region with the highest fatalities per 100,000 population.

\begin{figure}
    \centering
    \includegraphics[scale=0.53]{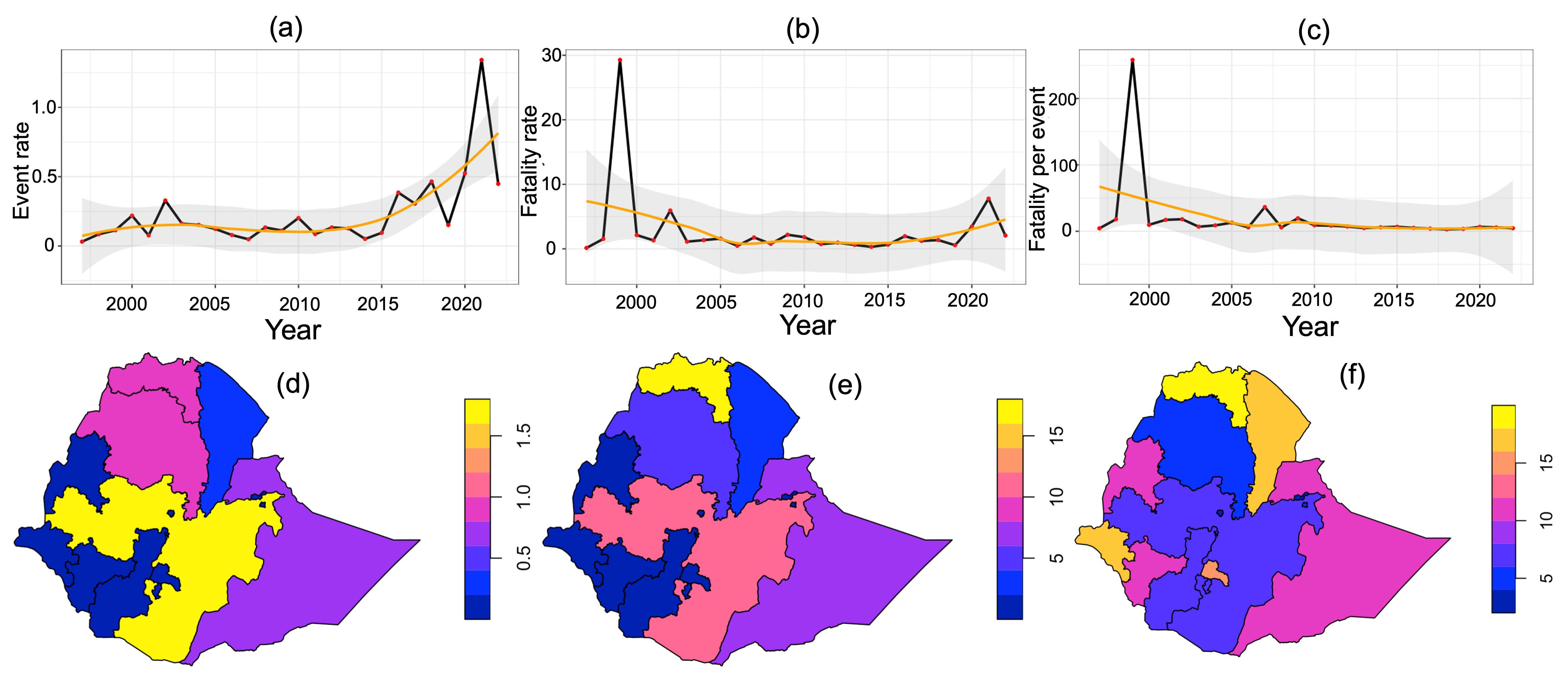}
   \caption{{(a) Violent event rate, (b) fatality rate per 100 000 population, and (c) fatality rate per violent event between the years 1997 and 2022, averaged for all regions. (d) Violent event rate, (e) fatality rate per 100 000 population, and (f) fatality rate per event for all regions of Ethiopia averaged for all years.} }
    \label{fig:spaceplot}
\end{figure}
\subsubsection*{Model inference}
 Table \ref{dic} presents the model adequacy for the count component using the structural form I. Given the estimated measures, the negative binomial distribution uniformly outperformed the other models since it has the least DIC and WAIC. Moreover, the negative binomial distribution has the highest estimated effective parameters. Table \ref{diclink2} shows the criteria for different structural forms adopted using the negative binomial distribution for the count component. {The table
also shows the adequacy criteria for the baseline structural form, which is a model without spatial and temporal terms in the count component}. Based on these criteria, the result shows there is a significant information gain using structural form II compared with form I and the {baseline model} since the DIC and WAIC are smaller for structural form II. This indicates that fatalities due to violent events exhibit significant spatial and temporal interactions. 
Hence,  for inference, the structural form II was adopted for both the binary and count components with corresponding Bernoulli and negative binomial response models. The sensitivity of the PC prior hyperparameters was investigated, and the most appropriate choice was adopted. The resulting graphs of the Conditional Predictive Ordinate (CPO) for the adopted hyperparameters are shown in Figure \ref{fig:cpo} in the Appendix. {Based on structural form II and negative Binomial distribution, the estimated threshold value was $c=0.99055$ (see Section 2 of the supplementary material), which is relatively high, indicating that a few numbers of zeros in the acquired data are zero counts.} Table \ref{hyper} shows the posterior point estimates and the $95\%$ credible intervals of the hyperparameters in the binary and count components. {The posterior mean of the spatio-temporal range parameter is $r=1.188$ in the binary component. With a distance of approximately 111 kilometers on land for one degree of latitude, this finding suggests that fatality occurrence at one location provides insight into the likelihood of occurrence at locations within $132\,km\, (1.188\times111km)$ radius. It also suggests that there is negligible spatial autocorrelation of fatality occurrence beyond $132\,km$. Additionally, the posterior mean of the spatial range is $r=1.555$ in the count component indicating that the spatial autocorrelation of fatality counts of two locations within $173\,km\, (1.555\times111km)$ radius is significant. These findings imply that fatalities that occur within locations less than $132km$ in the binary component and $173\,km$ in the count component are significantly correlated}. This result shows that the spatial correlation extent of fatality count is higher than the corresponding spatial extent of the fatality occurrence per event. That is, the likelihood of fatality counts is more similar in broader area coverage compared with the fatality occurrence. The posterior means of the spatio-temporal correlation are $\rho_\phi=0.574$ and $\rho_\phi=0.430$ for the binary and count components respectively. {These quantities imply that fatality occurrences across the country are positively correlated over the years, indicating that fatality in a given year is significantly informed by those that occurred in the previous year. }
The posterior means of the spatial marginal variance are $\sigma^2=1.104^2$ and $\sigma^2=0.972^2$ for the binary and count components respectively, which indicates almost the same spatial heterogeneity in the fatality occurrence and rates. 

\begin{table}[H]
\centering
\caption{Measure of model adequacy for the count component using structural form I.}
\begin{tabular}{lcccc}
  \hline
Model & DIC&WAIC&Effective Parameter \\ 
  \hline
  Poisson & 48340.98&79886.18 & -18624.14 \\ 
     \rowcolor{Gray}
  Negative Binomial  &23542.06&23704.61 & 379.45 \\ 
  Generalized Poisson &24315.35&24292.53&73.17\\
   \hline
\end{tabular}
\label{dic}
\end{table}

\begin{table}[H]
\centering
\caption{Measure of model adequacy for the linear predictor using the negative Binomial distribution for the count component.}
\begin{tabular}{lcccc}
  \hline
Structural form & DIC&WAIC&Effective Parameter \\ 
  \hline
{I}\,\,\,\,[Eq. \eqref{LinearPred1}] & 23542.06&23704.61 & 379.45 \\ 
     \rowcolor{Gray}
{II}\,[Eq. \eqref{LinearPred2}] &22975.55&22844.68 & 724.85 \\ 
Baseline*& 25233.11&25379.58 &22.97\\
   \hline
\end{tabular}\\
\footnotesize{$^*$ Structural form without spatial and temporal terms in the count component.}
\label{diclink2}
\end{table}

\begin{table}[H]
\centering
\caption{Posterior estimate of the model's hyper-parameters.}
\begin{tabular}{lrrrcccc}
  \hline
&\multicolumn{3}{c}{Binary component}&&\multicolumn{3}{c}{Count component}\\
  \cmidrule{2-4}\cmidrule{6-8}
 & Mean & 2.5\% & 97.5\% &&Mean & 2.5\% & 97.5\% \\
 \hline
$r$ & 1.188 & 0.868 & 1.563&& 1.555 & 1.228 & 1.940  \\ 
   \rowcolor{Gray}
$\rho_\phi$ & 0.574 & 0.435 & 0.698 && 0.430 & 0.300 & 0.556\\ 
$\sigma$ & 1.104 & 0.977 & 1.248  && 0.972 & 0.904 & 1.044 \\ 
 $\xi$&&&&& 1.567 & 1.469& 1.674 \\
 \hline
\end{tabular}
\label{hyper}
\end{table}

Table \ref{temp} shows {the statistics of the marginal posterior distribution of $\boldsymbol\beta$} for the effects of temperate season on fatality occurrence and rates. The temperate season was split into four seasons according to the northern hemisphere. That is, the Autumn season used as the base category comprises the months of September, October, and November. The effects of the Winter season (December, January, and February), Spring (March, April, and May), and Summer (June, July, and August) with respect to the base category are shown in the Table. The noninclusion of zero in the credible interval was used to determine the significance of the estimates. The result shows that fatality occurrence has a significant {odd ratio of}  $0.198\,=\, (1-e^{-0.221})$  for Winter and $0.196\,=\, (1-e^{-0.218})$ for Spring, {which indicates a lower likelihood of fatality occurrence in Winter and Spring compared with Autumn season respectively}. However, fatality occurrence is not significantly different during Summer compared with Autumn. Turning attention to the count component, the fatality rate per event is significantly higher in Summer compared with the base category, Autumn. However, the rate is not significantly different for Spring and Winter compared with Autumn.
\begin{table}[ht]
\centering
\caption{Effect of temperate season on fatality.}
\begin{tabular}{lccccccc}
  \hline
     & \multicolumn{3}{c}{Binary}  && \multicolumn{3}{c}{Count}  \\ 
     \cmidrule{2-4}\cmidrule{6-8}
 & Mean & 2.5\% & 97.5\%&& Mean & 2.5\% & 97.5\% \\ 
  \hline
  Autumn & 0.000 &  && &  &  &  \\ 
   \rowcolor{Gray}
  Winter & -0.221 & -0.418 & -0.025 && -0.023 & -0.135 & 0.088 \\
  Spring & -0.218 & -0.429 & -0.007 && 0.060 & -0.056 & 0.176 \\ 
    \rowcolor{Gray}
  Summer & -0.092& -0.296 & 0.112 && 0.135 & 0.026 & 0.245 \\ 
  \hline
\end{tabular}
\label{temp}
\end{table}


Figure  \ref{figsocial} shows the effect of social and armed groups on fatality occurrence and fatality rate. {The Figure shows the statistics of the marginal posterior distribution of $\boldsymbol\beta$ corresponding to the group covariate}. The thick lines are the posterior mean of these effects, and the fainted dashed lines are the corresponding 95\% credible intervals. The x-axis of the graph shows the ranking of the groups. Lower rank signifies a group associated with higher fatality. For emphasis, a higher effect for the binary component implies higher odds of fatality 
 occurrence, while a higher effect for the count component implies a high fatality rate per event. From the Figure, \textit{Ginbot7} ranked top in the binary component and lowest in the count component. This implies that events associated with the Ginbot7 group result in the highest likelihood of fatality occurrence but with the least fatality rate per event. In other words, events related to \textit{Ginbot7} rarely occur, but when it does, a high number of fatalities is expected. A comparable but positioned in the alternate form is the \textit{EDF} group. The events related to \textit{EDF} have the least likelihood of fatality occurrence but have a high likelihood of being linked to events that lead to high fatality per event.  In the binary component, the groups following $Ginbot7$ are Eritrea Army, SPLA, TPDM, OLA, ONLF, EUFF, TPLF, and EDF in increasing ranking position. In the count component, the groups following the \textit{Eritrea Army} are SPLA, TPDM, EDF, ONLF, OLA, TPLE, EUFF, and Ginbot7 in increasing ranking position. The most dreaded events are those events associated with groups whose effects in binary and count components fall above the null effect, $0$. In this case, it implies that fatality is not only frequent but also on a large scale. Examples include the events associated with the Eritrea army, ONLF and SPLA.
\begin{figure}[]
    \centering
\includegraphics[scale=.8]{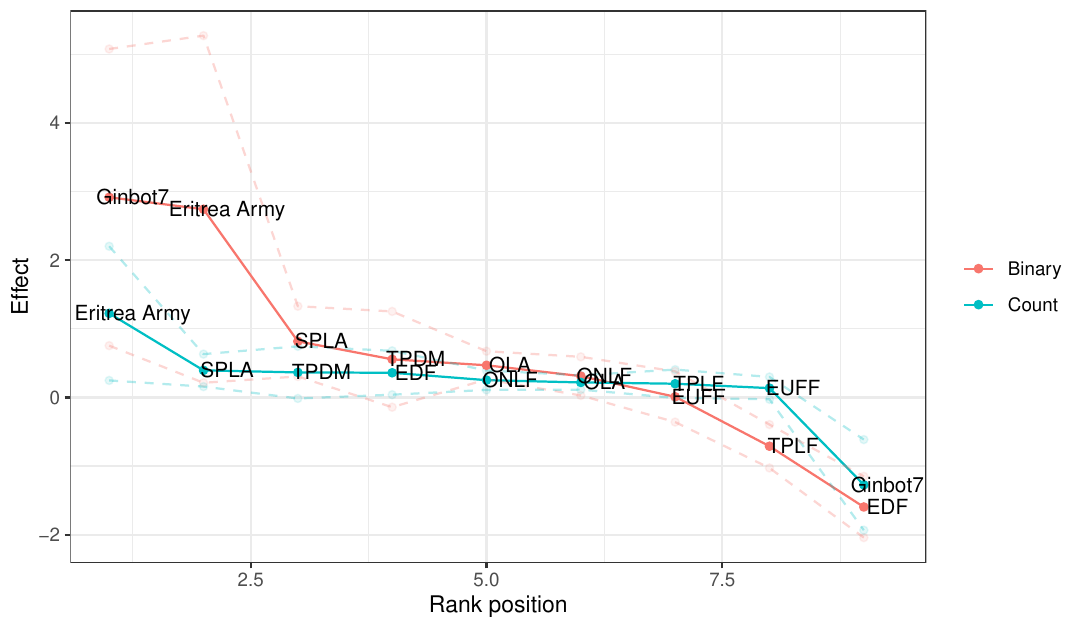}
\caption{Effect of the social, political, and armed group.}
    \label{figsocial}
\end{figure}

Figure \ref{figevent} presents the posterior estimates of the effect of event types on fatality. {It shows the statistics of the marginal posterior distribution of $\boldsymbol\beta$ corresponding to the event type covariate}. The thick dots denote the effect of the event types that are significant, while the fainted dots denote the effects that are not significant at 95\% credible intervals. The higher the effect, the higher the likelihood of fatality occurrence for the binary component and fatality rates for the count component. The complete descriptions of these event types can the found in the ACLED \href{https://acleddata.com/acleddatanew/wp-content/uploads/2021/11/ACLED_Codebook_v1_January-2021.pdf}{code book}. The result shows that events classified as \textit{Attack} have the highest likelihood of fatality
\begin{figure}[H]
    \centering
\includegraphics[scale=.75]{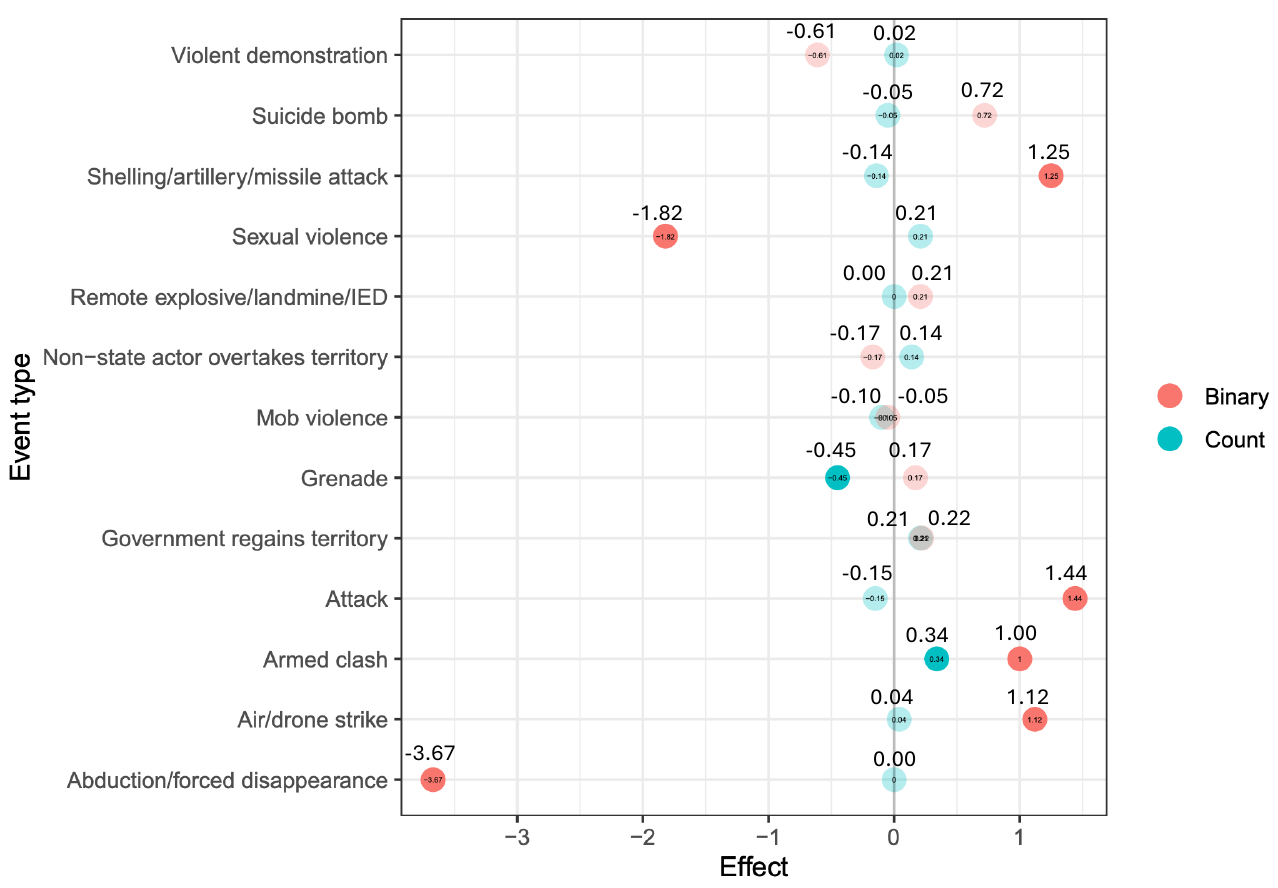}
\caption{Effect of the event types.}
    \label{figevent}
\end{figure}
\noindent occurrence while the events classified as \textit{Abduction/force disappearance} have the least likelihood of fatality occurrence. Moreover, events resulting in a high likelihood of fatality occurrence also include \textit{Shelling/artillery/missile attack, Armed clash, and Air/drone strike}, while the event that leads to the lower likelihood of fatality occurrence also include \textit{Sexual violence}. Turning attention to the count component, events classified as \textit{Armed clash} have a significantly higher fatality per event, while events classified as \textit{Grenade} have a significantly lower fatality per event. Unsurprisingly, Armed Clash is significant in both binary and count components. This result means Armed Clash event type simultaneously leads to frequent fatality occurrence and high fatality per event.

Figure \ref{spatemp1} and \ref{spatemp2} present the {projection of the estimated} spatio-temporal effects {across the entire region under study} for the binary and count components of the model respectively. The {figures show the
statistics of the marginal posterior distribution of $\boldsymbol\phi$.} 
In Figure \ref{spatemp1}, there was a steady increase in the burden intensity of fatality occurrence among the locations suffering from frequent fatality due to violent events from the year 1997 to 2004. The burden was severe among locations in Tigray, Amhara, Oromia, Gambela, and Somali regions over these years. Each of the listed regions took a fair share of the burden at one point during these years. For instance, in 2000, Tigray had the highest fatality occurrence, while in 2003, Somali had the highest fatality occurrence. In 2006, the burden dropped throughout the country but became worse in the following year (2007). The overall burden dropped again in the year 2008 but picked up in 2009. Somali bore the highest burden of fatality occurrence within these years of rapid changes. From 2009, the pattern increased consistently until 2019. Over the years,  locations at the border shared between Somali and Oromia consistently had a high burden of fatality occurrence in addition to Amhara and Afar. From 2020 to 2022 the burden slightly decreased, and locations in Oromia, Benshangul-Gumaz, and southwest Somali had the highest burden of fatality occurrence in the country.

Figure \ref{spatemp2} shows the posterior estimate of the spatio-temporal pattern in the count component. The fatality per event increased from the year 1997 to 1999 when it attained its peak throughout the study period. In these years, Afar, Tigray, in the north, and Oromia in the south had the highest burden of fatality per event. The burden dropped quickly in the following years. Between 2002 and 2005, Gambela, South West Ethiopia Peoples' Region, and SNNPR had the highest fatality per event. In 2006, Tigray had the highest burden, and locations on the border shared between Oromia and Amhara had the highest burden in the following year (2007). The fatality per event decreased afterward until 2019, and few locations in the county suffered from high fatality per event. In 2020, the high burden resurfaced and spans till 2022. In these years, high fatality per event was severe in locations found in Tigray and Benshangul-Gumaz.

\begin{figure}[H]
    \centering
     \includegraphics[scale=0.5]{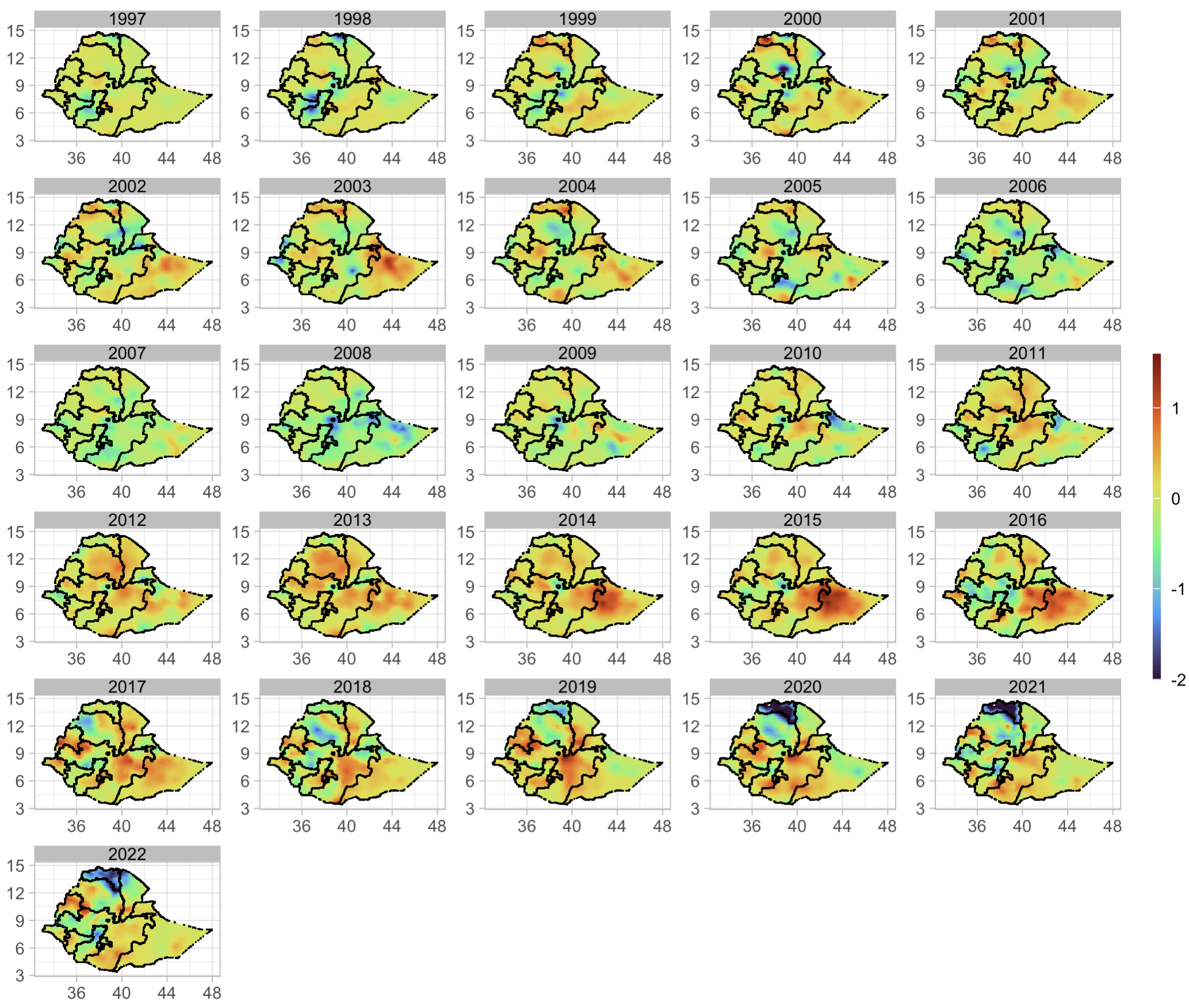}
    \caption{Projected posterior mean of the spatio-temporal effect in the binary component.}
    \label{spatemp1}
\end{figure}
\begin{figure}[H]
    \centering
    \includegraphics[scale=0.5]{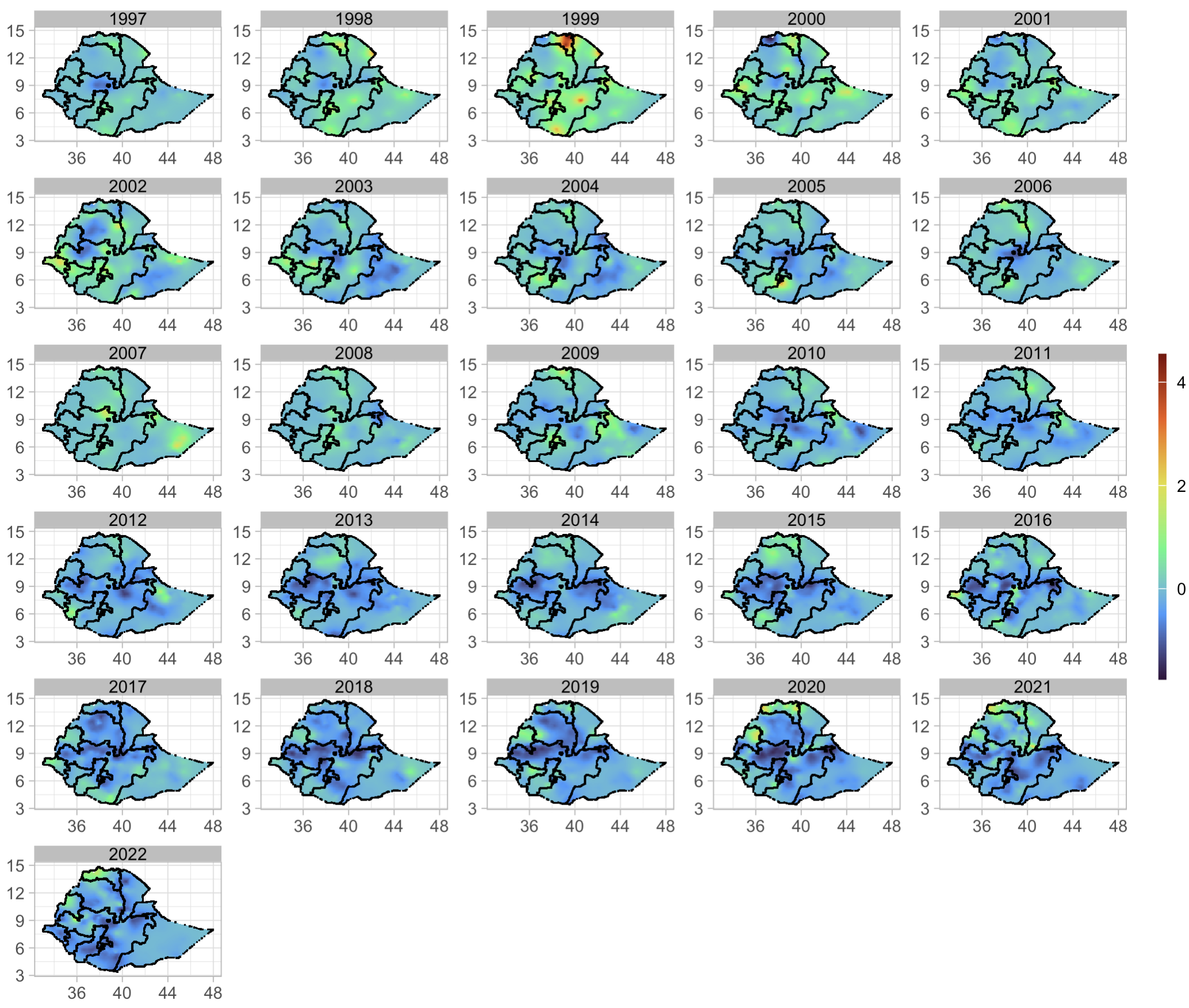}
    \caption{Projected posterior mean of the spatio-temporal effect in the count component.}
    \label{spatemp2}
\end{figure}

Figure \ref{uncertainty} shows the uncertainty map computed using Equation \eqref{postpred} and aggregated by administrative regions. The binary component presents the probability of fatality occurrence due to violent events, while the count component presents the probability of more than $20$ deaths per violent event. The result shows that the probability of fatality occurrence is higher in the southern region of the country and lower in the north. Specifically, five of the thirteen administrative regions had above 0.7 chance of fatality occurrence. These regions include Benishangul Gumz, Gambela, Oromia, Somali, and South West Ethiopia Peoples' Region. 
Moreover, the probability is highest in Benishangul-Gumuz and lowest in Tigray. However, the probability of more than 20 deaths per violent event is higher in the north and lower in the southern region of the country. Tigray administrative region had the highest probability ($0.558$) of more than $20$ deaths per violent event. The Benishangul-Gumuz region that follows had a probability of $0.306$. Benishangul-Gumuz had relatively high probabilities in both the binary and count components. The uncertainty of fatality over the years dropped on average, which corroborates the results obtained for the spatio-temporal patterns. For example, the probability of more than $20$ deaths in the last six years ($2017-2022$) of the study period is $0.105,\,0.0829,\,0.0928,\,0.413,\,0.367,$ and $0.148$ respectively.  
\begin{figure}[H]
    \centering
    \includegraphics[scale=0.7]{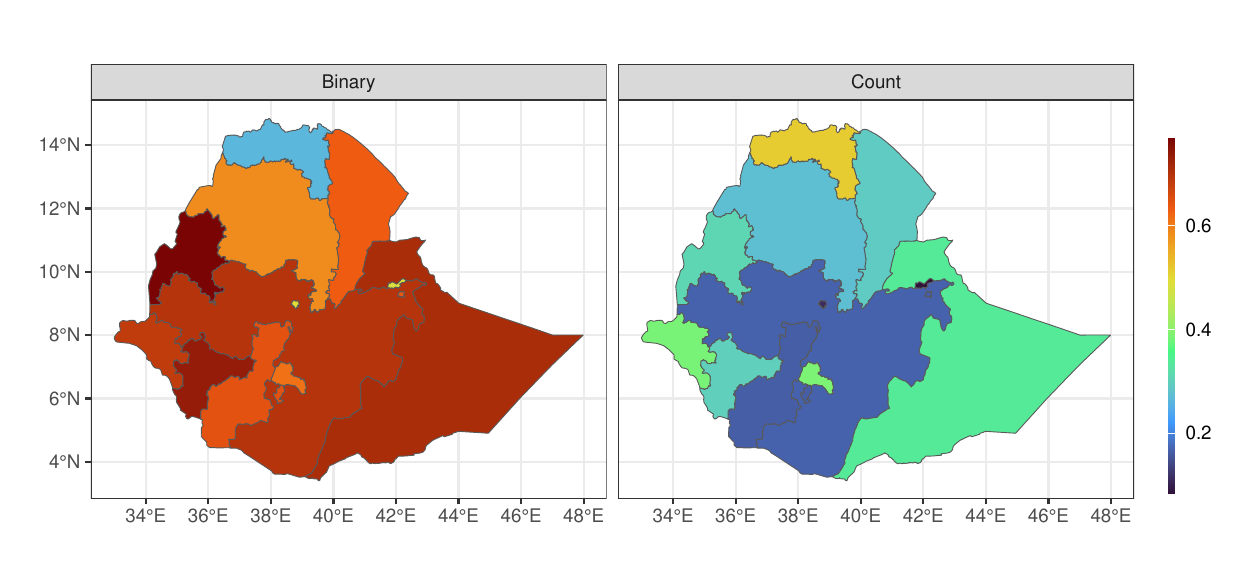}
    \caption{Uncertainty measure.}
    \label{uncertainty}
\end{figure}
\section{Discussion}\label{discus}
This work developed a two-part zero-inflated Bayesian generalized mixed model to unveil the spatio-temporal pattern of violent event-related fatality in Ethiopia. One of the components (binary) was used to model the occurrence or nonoccurrence of fatality, and the other (count) models the fatalities per violent event. The binary component was used to target fatality occurrence due to violent events by disregarding whether or not there are multiple fatalities for each event, while the count component targets the number of fatalities on the condition that a violent event had occurred. The SPDE was used to model the spatial dependence for each model component, and the \texttt{R-INLA} was used for estimation and projection. The result shows that the two model components are governed by different significant risk factors and exhibit different spatial and spatio-temporal patterns. These risk factors include events associated with top social, political, and armed groups in Ethiopia, the temperate season, and the type of event that took place.

Findings revealed that fatality occurrence and the associated fatality per violent event were lower during winter, while they were relatively enormous during summer. This finding supports the popular theory that law-breaking activities tend to slow down during the cold season and are projected to increase during the hot season. This finding corroborates the works of \cite{breetzke2012seasonal},  \cite{cohen44m}, and \cite{egbon2025spatial1}. This could be linked to the differences in social attitudes during these seasons. That is, the cold season tends to lower the likelihood of potential victims coming into contact with offenders, whereas, in the hot season, the likelihood is highest as potential victims are more likely to socialize and encounter violent offenders.
According to the general aggression model \citep{anderson1995hot},
an increase in temperature creates irritability and therefore violent actors are disposed to commit crimes. Moreover, in conflict regions, cold weather, rain, and mud make it harder to move large vehicles, tanks, and heavy weapons, thus lowering armed combat and consequently the fatality during this period. Therefore, armed state actors could capitalize on the weakness of violent offenders during winter and establish policies and strategies ahead of the summer, especially in conflict regions.

Findings show a significant impact of social, political, and armed groups on fatality in the country. While some categories of violent events associated with certain armed groups frequently lead to fatality occurrence, others that occur less frequently lead to high fatalities per event. Examples of such groups are Ginbot7, EDF, and TPLF. Moreover, findings also show that there exist certain groups that are both associated with frequent fatality and high fatality per violent event. Examples of these armed groups include the Eritrea army or its liberation front, SPLA, and ONLF. 
These findings could be linked to the Eritrean-Ethiopian war that took place between the years 1998 and 2000 \citep{murphy2016eritrean}, and the follow-up border conflict between the two countries for decades. 
Moreover, { the OLA and ONLF groups in Oromia and Somali were known to have demanded autonomy in these regions of Ethiopia, which gave rise to the
 historical conflict between Ethiopia's defense forces and the groups' militias. The conflict led to several damages causing displacement and fatalities, which led to the declaration of these groups as terrorist organizations by the government. The Ginbot7 militants also constituted a serious military threat to the Ethiopian government. The conflict spanned more than two decades until the groups declared a unilateral cease-fire in 2018 and were delisted from the government's terrorist list \citep{abdullahi2007ogaden,emirru2020multinational}}. 
    { The Ethiopian government initiated military operations against the TPLF in Tigray in November 2020. The conflict blew up beyond proportion until the Ethiopian government designated the TPLF a terrorist group} \citep{workneh2019counter,tegegne2021standardized,pellet2021understanding}, as activities of the TPFL militia have been linked to multiple fatalities in Mai Kadara, Galicoma, and Chena among other regions in Ethiopia \citep{tegegne2021standardized}. 
    Extremists have capitalized on the high ethnical heterogeneity among Ethiopians to operate and fulfill their goals in the region.
The conflict of the social, political, and armed groups against government forces brought more damage than good to the common Ethiopian people. Many are displaced from their homes and lack good healthcare due to the disruption of already weakened healthcare systems and the prevalence of severe food insecurity. This study recommends the Government, local authorities, and traditional rulers issue orders to their commanders to end indiscriminate attacks targeted at civilians and ensure that independent humanitarian organizations have complete access to civilians affected by conflicts. 

This work found disparities in fatality based on the event types that took place. In the event of an offensive Attack, fatalities occurrence is relatively at peak. According to ACLED, events categorized as Attacks are those attacks against civilians by armed actors \citep{acled2019armed}. This involves a deliberate or haphazard shooting at civilian targets that are of no military threat in order to kill, injure, and demoralize rival civilian populations.  This finding implies that the frequency of attacks on civilians and the consequent fatality occurrence is relatively the highest among all the event types. There have been reports of such attacks on civilians in different parts of Ethiopia, including the Tigray region, by different armed groups including military forces \citep{pellet2021understanding,yeneabat2020fighting}. The armed actors often use it as a punishment against their rival groups.
As expected, Shelling/artillery/missile, Armed clash, and Air/drone strikes were also found to have significantly higher odds of leading to fatality occurrence. 
However, findings reveal that sexual violence and abduction/forced disappearance rarely lead to fatality occurrence but demoralize victims. Furthermore, Armed clash events were revealed to lead to the highest number of fatalities per event in Ethiopia. Armed clashes are reported to be prevalent in Ethiopia as different social, political, and armed groups with conflicting interests are involved in armed fights, resulting in numerous deaths, including civilians.
Hence, this finding brings to the notice of the Ethiopian populace and government the variation in fatalities due to these event types to guide informed decisions, policing, and the resuscitation of economic and security policies that are appropriate for reducing arm inflow, conflict risks, and overcoming state fragility. Since these events are fueled by illicit inflow and movement of arms and ammunition, intercepting the flow at strategic borders could lower the occurrence rate and impede the strength of armed non-state actors \citep{yared2021conflict}.

It was found that there were no significant changes in the temporal pattern of fatality occurrence in the country over the study period. This implies that on average, the attitude of the populace towards conflict has hardly changed since 1997 till date. This may be attributed to the diversity in ethnicity and the so-called unfavorable governmental policy choices, and marginalization. Ethnic sentiment by members of a certain ethnic origin who consistently compare their position with other ethnic groups based on governmental policies develops into rebellion and conflicts \citep{yeshiwas2018transactive}. However, the findings showed that the number of fatalities per event on average declined consistently since 1997. The peak in fatality per event experienced in 1999 may be linked to the Eritrea-Ethiopian war \citep{murphy2016eritrean}, while the sudden rise in the fatality rate in 2007 can be attributed to the conflict between the Ethiopian army and the ONLF armed group in Somali region \citep{lyons2009ethiopia}. 

After adjusting for other effects, findings revealed that fatality occurrence was most prevalent in Somali, Oromia, and Benishangul-Gumuz regions, while a high number of fatalities per event was prevalent in the Tigray and Benishangul-Gumuz regions. Tigray region, for example, had low odds of fatality occurrence; however, they suffer a relatively severe fatality rate each time there is conflict.  Contrarily, the Somali and Oromia regions experience frequent fatality occurrences but relatively low fatality rates whenever there is conflict. Moreover, the Benishangul-Gumuz region experienced both frequent fatality occurrences and severe fatality rates. The high level of ethical diversity in the Benishangul-Gumuz region may be attributed to the fatality pattern given that previous studies have established a strong role of ethnic diversity in conflicts \citep{weidmann2011violence,esteban2012ethnicity}. Findings from the spatio-temporal pattern corroborate those obtained from the temporal fatality pattern. There is hardly a unidirectional spatio-temporal pattern of fatality occurrence. That is, the odds of fatality occurrence were relatively low in the first decade into the study, high in the second, and again low thereafter. However, the spatio-temporal pattern of the fatality rate became less severe as the year increased. 

The main limitation of this work is that the interaction between the two components of the model was not incorporated. This can be achieved by simply allowing the main model effects to interact through the prior distribution. The covariance matrices of both components' priors could be linked through a correlation parameter. However, such model improvement introduces computational complexity that may require high computational power. Future work can investigate whether the incorporation of interaction significantly improves the model relative to the computational challenges. Another limitation is probably fatality count {misreporting}. Future work can incorporate statistical methods commonly used to correct such inconsistencies in the data. Future work can also conduct an in-depth study to quantify the severity of fatalities linked to popular armed actors at the sub-national level, which could be informative for strengthening local policies against violent activities in the country. {Additional limitation is the assumption that the locations where fatalities occur are not random. In other words, it assumes that the location where a fatality occurs is predetermined before actually observing it. To address this limitation, future research could explore a spatial point pattern model that can account for the location randomness observed in the data. {Moreover, the small number of covariates included in the data analysis {could underestimate the probability of fatality per event}. Future work can improve on the data analysis by considering a larger set of covariates.}


\section{Conclusion}\label{concl}
    Ethiopia has experienced several armed conflicts that have led to the displacement of citizens and caused preventable fatalities. As shown from the findings, social, political, and armed agitators have earnestly played a key role in the unrest. The developed maps reveal a comparative fatality burden over the years, which could be matched with previous governmental policies and local economic performance to understand and plan for the future. Despite there being a relative decline in fatality per violent event, the rate is still high. The government, policymakers, political elites, and traditional or religious rulers must then exercise informed, decisive steps to end human rights violations. This is achievable by issuing unambiguous public orders to the military commanders, regional police, and local armed groups to cease indiscriminate and extrajudicial killings, rape, and looting. State actors should be trained and retrained on ways to deal with civilians. Stakeholders must seek international support for the affected populace and region. It is important that the government engage in nutritional assessment of the affected region to mitigate under-five deaths due to malnutrition.

\newpage
\bibliography{references}
\appendix
\section*{Appendix}

\setcounter{figure}{0} \renewcommand{\thefigure}{A.\arabic{figure}}
\setcounter{table}{0} \renewcommand{\thetable}{A.\arabic{table}}
 \begin{figure}[H]
     \centering
  \subfloat[]{  \includegraphics[scale=0.6]{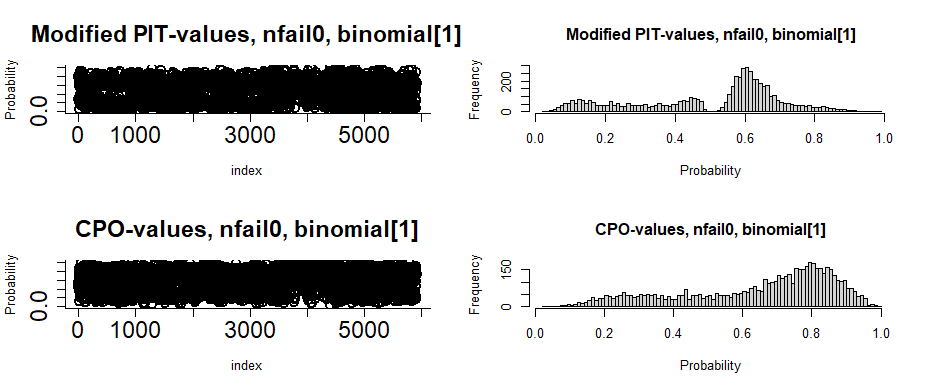}}\\
  \subfloat[]{ \includegraphics[scale=0.6]{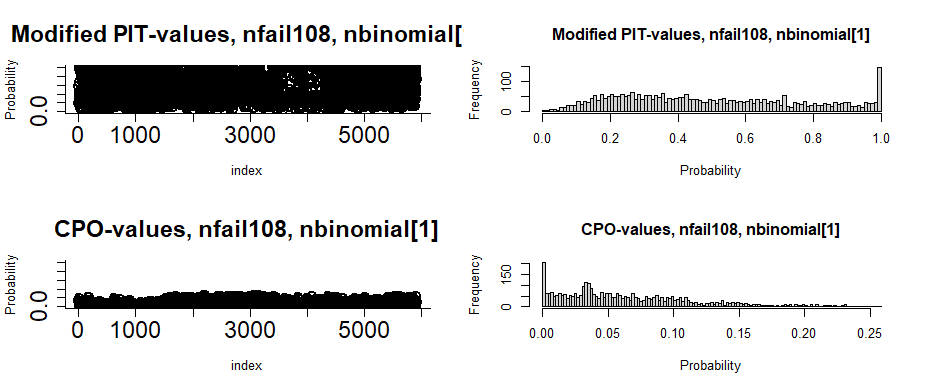}}     
     \caption{The Conditional predictive ordinate and Probability-integral transform values.}
     \label{fig:cpo}
 \end{figure}
 
\end{document}